\documentclass[aps,a4paper,floatfix, nofootinbib]{revtex4}

\usepackage[utf8]{inputenc}

\usepackage{graphicx,color}
\usepackage{amsmath,amssymb,amsfonts,amsthm,amscd,bm}
\usepackage{booktabs}
\usepackage{cancel}

\def\tilde{\widetilde}
\def\bar{\overline}
\def\hat{\widehat}
\def\*{\star}
\def\[{\left[}
\def\]{\right]}
\def\({\left(}      

\def\){\right)}

\def\zbar{{\bar{z} }}
\def\frac#1#2{\dfrac{#1}{#2}}
\def\inv#1{\dfrac{1}{#1}}

\def\d{\partial}

\def\2pi{\hbox{$2\pi i$}}

\def\dsl{\raise.15ex\hbox{/}\kern-.57em\partial}
\def\Dsl{\,\raise.15ex\hbox{/}\mkern-.13.5mu D}
\def\vep{\varepsilon}

\def\vphi{\varphi}
      \def\CF{{\cal F}}

   \def\CN{{\cal N}}   \def\CO{{\cal O}}
      
\def\CS{{\cal S}}

\def\2pi{\hbox{$2\pi i$}}

\def\dsl{\raise.15ex\hbox{/}\kern-.57em\partial}
\def\Dsl{\,\raise.15ex\hbox{/}\mkern-.13.5mu D}
\font\numbers=cmss12
\font\upright=cmu10 scaled\magstep1
\def\stroke{\vrule height8pt width0.4pt depth-0.1pt}
\def\topfleck{\vrule height8pt width0.5pt depth-5.9pt}
\def\botfleck{\vrule height2pt width0.5pt depth0.1pt}
\def\Zmath{\vcenter{\hbox{\numbers\rlap{\rlap{Z}\kern
    0.8pt\topfleck}\kern 2.2pt
    \rlap Z\kern 6pt\botfleck\kern 1pt}}}
\def\Qmath{
    \vcenter{\hbox{\upright\rlap{\rlap{Q}\kern3.8pt\stroke}\phantom{Q}}}}
\def\Nmath{\vcenter{\hbox{\upright\rlap{I}\kern 1.7pt N}}}
\def\Cmath{\vcenter{\hbox{\upright\rlap{\rlap{C}\kern
                   3.8pt\stroke}\phantom{C}}}}
\def\Rmath{\vcenter{\hbox{\upright\rlap{I}\kern 1.7pt R}}}
\def\Z{\ifmmode\Zmath\else$\Zmath$\fi}
\def\Q{\ifmmode\Qmath\else$\Qmath$\fi}
\def\N{\ifmmode\Nmath\else$\Nmath$\fi}
\def\C{\ifmmode\Cmath\else$\Cmath$\fi}
\def\R{\ifmmode\Rmath\else$\Rmath$\fi}
\def\barray{\begin{eqnarray}}
\def\earray{\end{eqnarray}}
\def\beq{\begin{equation}}
\def\eeq{\end{equation}}

\def\n{\noindent}

\def\vphi{\varphi}
\def\vphibar{\bar{\varphi}}

\def\Li{{\rm Li}}

\def\AA{\leavevmode\setbox0=\hbox{h}
\dimen0=\ht0 \advance\dimen0 by-1ex\rlap{\raise.67\dimen0\hbox{\char'27}}A}

\def\Li{{\rm Li}}

\makeatletter
\def\iddots{\mathinner{\mkern1mu\raise\p@
\vbox{\kern7\p@\hbox{.}}\mkern2mu
\raise4\p@\hbox{.}\mkern2mu\raise7\p@\hbox{.}\mkern1mu}}
\makeatother

\def\Li{{\rm Li}}

\def\Tbar{\bar{T}}
\def\vtheta{\vartheta}
\def\vep{\varepsilon}
\def\shG{{\rm shG}}
\def\cdd{{\rm cdd}}
\def\mr{\tfrac{mR}{2}}
\def\dim#1{ {\bf [[} #1 {\bf ]]}  }

\theoremstyle{plain}

\theoremstyle{remark}

\begin{document}

\def\cstarUV{c^*_{UV}}

\title{
Thermodynamics of  $T \Tbar$ perturbations of  some  single particle field theories
}
\author{
 Andr\'e  LeClair\footnote{andre.leclair@gmail.com}
}
\affiliation{Cornell University, Physics Department, Ithaca, NY 14850} 

\begin{abstract}

We study the  Thermodynamic Bethe Ansatz  (TBA) equations for  pure $T\Tbar$ perturbations of some simple integrable  quantum field theories with a single bosonic or fermionic particle,  in particular the 
massive sinh-Gordon model and its ultraviolet (UV) limit which is a deformation of the conformally invariant  free massless boson.  
Whereas the TBA equations for $T\Tbar$ deformations of massive theories are in principle known,  the TBA equations we propose  for the deformations of conformal field theories (CFT's) are relatively new and require a special factorization in rapidity variables of the CDD factor for the scattering of the massless particles.  
The latter TBA equations can be solved exactly and reproduce the known results for the ground state energy on a cylinder of circumference $R$ which were previously obtained using different methods based for instance on  the Burgers differential equation.   
Special attention is paid to the c-theorem in this context which is discussed in some detail.    
     For positive infra-red (IR) central charge $c_{IR}$,  for flows consistent with the c-theorem the ground state energy develops a  (previously known)  square-root singularity towards the UV,  which strongly suggests the theories are UV incomplete in these physically important cases.      
We suggest that the singularity indicates a tachyonic vacuum instability.  Other cases  with $c_{IR} < 0$ do not have this singularity are interpreted as being UV complete with 
$c_{UV} = 0$.  
We extend our results to a continuously variable $c_{IR}$ by introducing a chemical potential and suggest this as a possible toy model 
 for the $T\Tbar$ perturbed Liouville theory.

\end{abstract}

\maketitle
\tableofcontents

\section{Introduction}

Recently there has been much interest in so called $T\Tbar$ perturbations of 2 dimensional quantum field theory
\cite{ZTT, Tateo1, SmirnovZ, Verlinde,  Dubovsky, Tateo2, Rosenhaus}.   For a review and many additional references see for instance \cite{Jiang}.    
Every theory has a conserved stress-energy tensor, 
\beq
\label{Conv1}
T =-2 \pi \, T_{zz} , ~~~\Tbar = -2 \pi \, T_{\zbar\zbar}, ~~~ \Theta = 2 \pi\,  T_{z\zbar} , 
\eeq
where $z=x+i y$, $\zbar = x-i y$ are euclidean light-cone coordinates and $\Theta$ is its trace.     It was shown in \cite{ZTT} that the dimension 4 irrelevant operator 
\beq
\label{Conv2}
T \Tbar = 4 \pi^2 \( T_{zz} T_{\zbar\zbar} - (T_{z \zbar})^2 \) 
\eeq
is well defined.  One then considers a   theory defined by the action   
\beq
\label{Conv3}
 \CS =  \CS_0 +  \frac{\alpha}{\pi^2} \int d^2 x \, T\Tbar .
\eeq
More precisley:
\beq
\label{CS2}
\frac{d \CS}{d \alpha} = \inv{\pi^2} \int d^2 x \, \, (T\Tbar)_\alpha 
\eeq 
as an effective action.
$\CS_0$ is the unperturbed theory,  which could be massive or conformally invariant, and it could be  integrable or not.    That 
$T\Tbar$ is an irrelevant operator contributes to the interest in these models,  since it is unclear thus  far what kind of ultra-violet completion they may have,  if any. 
This is the main open question of this subject.     

Smirnov and Zamolodchikov showed that if $\CS_0$ is integrable,  then so is its $T\Tbar$ perturbation.
It was also shown that the effect of the $T\Tbar$ perturbation is to multiply the factorized S-matrix by a CDD factor.   

Another topic that has generated interest in these models is the connection to 2-dimensional gravity and string theory.,  for instance
Jackiw-Teitelboim (JT) gravity\cite{Jackiw,Teitel} and ${\rm AdS}_3/ {\rm CFT}_2$.         Namely,  the $T\Tbar$ perturbation may be 
viewed as a gravitational dressing of $\CS_0$ \cite{Dubovsky,Dubovsky2,Cardy,Tateo3,Verlinde,Hartman,Frolov, Oku}\footnote{This is just a partial list of references. For a review see
\cite{Jiang}.}.  
  Although a very interesting development,  this will not be the focus of this article, although some comments will be made in this regard.   

In this article we consider the Thermodynamic Bethe Ansatz (TBA) for $T\Tbar$ perturbations  of 
some relatively simple models involving a single bosonic or fermionic particle.  
The TBA in this context has not been studied in much detail in the existing literature.  
Early important works for the case of a free massless boson are \cite{DubovskyTBA,Tateo0} \footnote{We only became aware of these very early works after the first version of this article was written.}.     
 For a massive theory such as the sinh-Gordon model it is straightforward to 
incorporate the CDD factor by following the general construction in \cite{ZTBA,KM},  although this model has not been studied in any considerable detail. 
The main focus of this article are massless theories which are $T\Tbar$ deformations of CFT's.   
     The $T\Tbar$ perturbation of a free massless boson 
describes the UV limit of the perturbed sinh-Gordon model.    For this kind of massless theory, and also its fermionic analogs,  we will develop a novel type of TBA equation,  based on a generalization of  the
formalism of Zamolodchikov and Zamolodchikov for integrable massless scattering \cite{ZZ}.   A precursor to these equations can be found in \cite{DubovskyTBA,Tateo0}. 
The advantage of the formulation presented here is that one can more readily borrow results from massless flows \cite{ZZ,Fendley}.  
   As we will show,  the  novelty lies in the need to factorize the CDD factor in rapidity space.     
As we will show,  our proposed TBA equations can be solved exactly and reproduce known results  \cite{SmirnovZ,Tateo1} on the ground state energy on an infinite cylinder of radius $R$,  and this serves as an important check of our formalism since the methods based for instance on the Burgers equation are rather different.

   The TBA for more  conventional massless flows tracks the renormalization group  flow between
known UV and infra-red (IR) fixed points \cite{Fendley}, i.e. theories that are UV complete.        Whether $T\Tbar$ perturbations are UV complete was left
as an open question in \cite{SmirnovZ}.     Below we will use our proposed TBA to study this question, and   for some cases that appear to be UV complete  we will  propose values of the (effective) Virasoro central charge $c_{UV}$.
($c_{IR}$  is known a priori).     We will also discuss the c-theorem in this context since it provides some insights on the interpretation of the RG flows.      

The remainder of this paper is organized as follows.    Sections II, III  are brief reviews of well-known results  which we  build upon in subsequent sections.     Section II mainly serves to define conventions for the models we will discuss.   Section III reviews the TBA for the sinh-Gordon model;  this serves to introduce the main elements of the TBA,  and also leads naturally to the main features of massless TBA systems that we will build upon.    Although the sinh-Gordon model is not the main focus of this article,  it also serves to point out certain general issues regarding renormalization group flows for massless verses massive theories.      
In Section IV we present TBA equations for  the $T\Tbar$ perturbation of the massive sinh-Gordon model.    
As stated above,  our main focus is massless scattering which occurs in the UV limit,  hence we will not study further the TBA solutions for the massive sinh-Gordon case.  
   In the deep UV the theory is described by a $T\Tbar$ perturbation of a free massless boson.
The latter, and its analog for fermions,  is studied in Section V where we show that the massless TBA equations can be solved exactly under our proposed factorization prescription.  
   In Section VI we analyze the renormalization group flows based on the TBA equations for a more general class of models with variable $c_{IR}$ and discuss the c-theorem in this context.   In section VII we shown how to  obtain  the TBA equations for a continuous variable $c_{IR}$ by incorporating a chemical potential,  and suggest that this is a kind of toy model for 
   $T\Tbar$ deformation of the Liouville theory.

\section{The sinh-Gordon model and  related Liouville theory}

\subsection{Sinh-Gordon}

The sinh-Gordon field theory can be defined by the action
\beq
\label{shGS}
S = \int d^2 x \( \inv{8\pi} \d_\mu \phi  \, \d_\mu \phi + \lambda \, \cosh (b \phi)  \), 
\eeq
where $b$ is a dimensionless coupling and $\lambda$ sets a mass scale.   
We have chosen the normalization of the kinetic term such that for the free boson at $\lambda =0$,  one has $\phi(z, \zbar) = \vphi(z) + \vphibar (\zbar)$,   
with  two-point functions
\beq
\label{prop}
\langle \vphi (z) \vphi (w) \rangle = - \log (z-w),  ~~~~\langle \vphibar (\zbar ) \vphibar  (\bar{w} ) \rangle = - \log (\zbar - \bar{w}), 
\eeq
where  $z= x+i y$, $\zbar = x- iy$.  

Let  $\dim{\CO (x) }$ denote the scaling dimension in mass units of the operator $\CO (x)$.   In a conformal field theory, 
$\dim{\CO}= \Delta + \bar{\Delta}$ where $\Delta, \bar{\Delta}$ are the left, right scaling dimensions.   
One has 
\beq
\label{dim}
 \dim{  e^{b\phi} } = -b^2 \, .
\eeq
Thus the operator $\cosh b\phi$ is always relevant for real $b$.   The coupling $\lambda \propto m^{2 + b^2} $,  where $m$ is the mass of the fundamental particle.

The spectrum consists of a single particle of mass $m$,  and the energy and momentum of this particle can be parametrized as 
\beq
\label{rapid}
E = m \cosh \theta,~~~~~ p = m \sinh \theta.
\eeq
The two-particle S-matrix is known to be
\beq
\label{Smatrix} 
 S_{\rm shG} (\theta) = \frac{\sinh \theta - i \sin (\pi \gamma) }{\sinh \theta + i \sin (\pi \gamma)},  ~~~~~~~\gamma \equiv \frac{b^2}{2 + b^2}
\eeq
where $\theta = \theta_1 - \theta_2$.  
We point out that allthough the above S-matrix can be considered a ```CDD"  factor since it satisfies all the requirements,   by itself it has nothing to do with 
$T\Tbar$ deformations since the sinh-Gordon model is a relevant rather than irrelevant perturbation of a free boson; more on this below.  
Thus not all S-matrices built from pure CDD factors are related to $T\Tbar$ perturbations.       

Under the strong-weak coupling duality
\beq
\label{duality}
b \to \frac{2}{b}, 
\eeq
$\gamma \to 1- \gamma$ and the S-matrix is invariant.   Thus we only have to consider the range $0<b<\sqrt{2}$.

The ultraviolet  limit corresponds to $m\to 0$ since all energies are much greater than $m$ in this limit.   In this limit $\lambda \to 0$ and the theory becomes a free boson. 
The infrared  limit is on the other hand  $m \to \infty$.   The particle becomes infinitely heavy and decouples,  leaving an empty theory.  
Thus
\beq
\label{UVIR}
c_{IR} = 0, ~~~~~c_{UV} = 1
\eeq
where $c$ is the Virasoro central charge.

\subsection{Stress-energy tensor for the related Liouville field theory}

First,  we wish to stress that,   in spite of the similarity, there is no physical massless  limit of the sinh-Gordon model that gives the Liouville theory,   since the latter has a variable central charge. 
In this section we collect some well-known facts concerning the stress-energy tensor and central charge for the Liouville theory that we will need for the discussion in 
Section VII.
There are many more important results known about the Liouville field theory than the simple results presented here.  See for instance \cite{Teschner} and references therein.   

\bigskip

There are 3 separate regions to distinguish that we list below. 

\bigskip

\n 
{\bf {$25<c<\infty $.}} ~~
To define the theory,  following the conventions above for the sinh-Gordon model, we  start with the action
\beq
\label{LiouS}
S = \int d^2 x \( \inv{8\pi} \d_\mu \phi  \, \d_\mu \phi + \hat{\lambda}  \,  \exp ({b \phi} )  \). 
\eeq
With the above convention for the action,  the Liouville potential is ``half" of the sinh-Gordon potential.  
However this action does not completely define the theory since $e^{b \phi}$ has anomalous dimension that depends on $b$,  whereas the Liouville theory by definition 
 is conformally invariant. 
One needs to introduce a  ``background charge"  $a_0$ in order to make $e^{b\phi}$ exactly marginal,  i.e. of dimension $2$.   The stress tensor is now
\beq
\label{stress}
T_{zz} = T = - \inv{2} \d_z \vphi \d_z \vphi + a_0 \,  \d_z^2 \vphi
\eeq
and similarly for $\Tbar$.  One now has
\beq
\label{dima0}
\dim{ e^{a \phi} }= -a^2 + 2 a a_0 
\eeq
and the central charge becomes
\beq
\label{aa0}
c = 1 + 12 a_0^2.
\eeq
Setting $\dim{ e^{b \phi} } = 2$ determines $a_0$:
\beq
\label{a0}
a_0 = \inv{2} ( b + 2/b )
\eeq
and 
\beq
\label{cbc}
c = 1 + 3 (b + 2/b)^2  ~ \geq 25 .
\eeq
Note that $a_0$ and $c$ are invariant under the duality $b \to 2/b$, and the value $c=25$ occurs at the self-dual point $b=\sqrt{2}$. 

\bigskip\bigskip

\n {\bf {$-\infty<  c < 1 $}.} ~~
This regime can be viewed 
as  changing  to a time-like signature:  make the sign of the kinetic term in \eqref{LiouS}  negative, then rescale 
$\phi \to i \phi$ to restore the conventional two-point functions of $\phi$.     This amounts to $b\to i \beta$ with real $\beta$, 
and the above formulas can be analytically continued.    Letting also $a_0 \to i \alpha_0$,  one obtains
$\alpha_0 = (\beta - 2/\beta)/2$  and the central charge becomes 
\beq
\label{aa02}
c = 1 - 12 \alpha_0^2 = 1-3\( \beta - 2/\beta\)^2 \leq 1 .
\eeq
Note that  $\alpha_0$ and  $c$ are  invariant under the duality $\beta \to - 2/\beta$ ($c$ is also invariant under $\beta \to 2/\beta$).  The value $c=1$ occurs at $\beta = \pm  \sqrt{2}$.   

\bigskip\bigskip

\n {\bf {$1 < c< 25$ }.} ~~
This elusive region is important for non-critical  bosonic string theory with space-time dimension less than $26$,  since the world sheet central charge for matter plus Liouville 
must equal $26$.  The Liouville component comes from the gravitational sector of the 2 dimensional Polyakov world sheet action.  The complication arises since in order for $c$ to be in this range,  the coupling $b$ must be complex with both real and imaginary parts. 

\bigskip\bigskip
For the thermodynamic approach developed in Section VII we will be able to cover all three of the above regions in a uniform manner by introducing a chemical potential and varying the statistics from boson to fermion.

\section{TBA for the sinh-Gordon model and its massless limit}

In this section we review the  well-known Thermodynamic Bethe Ansatz,  since we will need to build on this formalism when generalizing it  in the subsequent sections
to deal with $T\Tbar$ deformations.  
The object of interest is the free energy density $\CF$ as a function of temperature $T$.   Since finite temperature field theory  amounts to putting the theory on
an infinite  cylinder with circumference $R=1/T$,  one can also view this  quantity as the ground state energy $E(R)$  of the model on a circle of circumference $R$.   
The relation is simply
\beq
\label{FE}
\CF (T) =  E(R) / R ,  ~~~~~~R=1/T .
\eeq
It is standard to express these quantities in terms of a scaling function $c(m/T)$ 
\beq
\label{cR}
\CF = - \frac{\pi}{6} \, c(m/T)\,  T^2 .
\eeq
where $m$ can be identified with the physical mass of the sinh-Gordon particle.  The normalization in \eqref{cR} is such that $c=1$ for a free massless boson.  
For a conformal theory,  $c(mR)$ is scale invariant, i.e. independent of $mR$.   For unitary theories it is  equal to the Virasoro central charge $c_{\rm vir}$,  otherwise it is
$c_{\rm vir} - 24 d$ where $d$ is the lowest scaling dimension of fields in the theory.    
The TBA allows a calculation of $c(mR)$ which can track the renormalization group (RG) flow.    

We introduce a quantum statistical parameter $s=\pm 1$,  where $s=1$ ($s=-1$) corresponds to bosons (fermions).   
When $\lambda=0$ in the sinh-Gordon model,  the theory is a free boson and the proper quantum statistical mechanics 
requires $s=1$ in order to obtain the correct $c=1$.    If one had chosen $s=-1$,  then $c=1/2$.    However interactions can change $s$.  
The criterion is $S(0) = s$.  For the sinh-Gordon model $s=-1$, i.e. fermionic,  and it is the interactions that give the correct $c_{UV} =1$ while treating the particles as
fermions.    
Generally speaking,  for interacting theories $s=-1$, however we continue to display $s$ for the  flexibility we will need in subsequent sections, since we will require bosonic
TBA systems in some cases.    The important issue of $s=-1$ verses $s=1$ was explored in detail in \cite{MussardoSimon},  where it was argued that bosonic type S-matrices are generally pathological.      What is more important about the latter work is that the kinds of square-rrot singularities encountered in $T\Tbar$ deformations were already observed there,
and it was recognized that this is related to perturbation by irrelevant operators.  
 However, based on some of our results below,  it  appears that irrelevant perturbations can provide counterexamples to the nearly universal $s=-1$.

\subsection{Massive case}

Introduce a so-called pseudo-energy $\vep (\theta )$ which is a solution to the integral equation:
\beq
\label{TBAmass}
\vep (\theta) = mR  \, \cosh \theta - \mu R +s G \star \log\( 1-s e^{-\vep} \)
\eeq
where $\mu$ is a chemical potential, which we will set to zero until Section VII. 
The kernel $G$ is 
\beq
\label{G}
G(\theta) = -i \d_\theta \log S(\theta) .
\eeq
Unitarity $S(\theta) S(-\theta) =1$ implies $G(-\theta) = G(\theta)$.
We have defined 
\beq
\label{convo}
\( G \star f \) (\theta) = \int_{-\infty}^\infty \frac{d\theta'}{2\pi} \, G(\theta - \theta') f(\theta') 
\eeq
for an arbitrary function $f(\theta)$.  
For the sinh-Gordon model, 
\beq
\label{GshG}
G_{\rm shG} (\theta)  = \frac{2 \cosh \theta\, \sin \pi \gamma} {\sinh^2 \theta + \sin^2 (\pi \gamma)} .
\eeq
Finally the scale dependent central charge is 
\beq
\label{cmass}
c(mR) = - \frac{3s}{\pi^2} \int_{-\infty}^\infty d\theta ~ (mR \cosh \theta) \, \log \( 1-s e^{- \vep (\theta)} \) .
\eeq

\subsection{Massless limit}
In the UV limit energies are much larger than $m$,  so this limit corresponds to $m\to 0$, namely massless particles.     
In order to determine the ultra-violet central charge $c_{UV}$, we take a massless limit of the above TBA following the framework developed in \cite{ZZ}.  
For right-movers we let $\theta = \theta_R +a$ and let $a\to \infty$ and $m\to 0$ keeping $m e^a$ fixed.   
For left-movers,  $\theta = \theta_L -a$  and we take the same limit.  
 The result is 
\barray
{\rm right ~ movers:} ~~~~~ E &=&  p= \tfrac{m}{2} e^{\theta_R} \cr
{\rm left  ~ movers:} ~~~~~ E &=& - p= \tfrac{m}{2} e^{-\theta_L}  .
\earray
Here $m$ is not the mass of a physical particle since $c$ is independent of $mR$ by the conformal invariance.  
The parameter $m$ is just  needed to give $E,p$ units of energy. 
The TBA equations have decoupled $L$ and $R$ pseudo-energies and now read 
\barray
\vep_R (\theta_R) &=&  \mr  e^{\theta_R} + s \,G_{RR} \star \log \( 1-s e^{-\vep_R (\theta)} \)  \cr
\vep_L (\theta_L) &=&  \mr  e^{- \theta_L} + s\,  G_{LL} \star \log \( 1-s e^{-\vep_L (\theta)} \)  
\label{TBAmassless}
\earray
where 
\beq
G_{RR} (\theta ) =  G_{LL} (\theta) = G_\shG  (\theta) .
\eeq
It is implicit that the convolution integrals involve $G_{RR} (\theta_R - \theta_R')$ and similarly for $G_{LL}$.  
Note that $G_{RR} (- \theta ) = G_{LL} (\theta)$, which implies 
\beq
\label{vepLR}
\vep_R (-\theta)    = \vep_L(\theta).
\eeq
Finally,
\beq
c(mR) = c_L (mR) + c_R (mR)
\eeq
where 
\barray
c_R &=&  - \frac{3s}{ \pi^2} \,  \int_{-\infty}^\infty d\theta_R \, \( \tfrac{mR}{2} \, e^{\theta_R} \) \, \log\( 1-s\, e^{-\vep_R (\theta_R) } \)  \cr 
c_L &=& -  \frac{3s}{ \pi^2} \,  \int_{-\infty}^\infty d\theta_L  \, \( \tfrac{mR}{2} \,  e^{- \theta_L} \) \, \log\( 1-s\, e^{-\vep_L (\theta_L) } \)  .
\label{cLR}
\earray

As already stated, for the above TBA system $m$ is an unphysical parameter since the TBA  describes a conformal field theory. 
The independence of $c$ on $mR$  can be seen by
shifting $\theta_{R,L}$ to remove the $mR$ dependence of equations \eqref{TBAmassless}.     Note also that $c_L = c_R$. 

\subsection{Extracting the central charge from the TBA}

\def\htilde{k}

Determining $c_{UV}$ from the above TBA is a standard computation which we now sketch.   Details can be found in \cite{ZTBA,KM}.  
We will not use the following results in subsequent sections;   we include it mainly to show that the UV central charge is equal to $1$ independent of $b$.
  In the cases of interest thus far,
$\vep_R = \vep_L 
\equiv \vep$.    It can be shown that up to some energy scale $\sim 1/R$,   $\vep$ is a constant $\vep_0$.     This constant is a solution to the equation 
\beq
\label{trans}
\vep_0 = s\, \htilde \,\log \( 1-s e^{- \vep_0} \),  ~~~~~ {\rm with}  ~~~~~\htilde = \int_{-\infty}^\infty \frac{d \theta}{2 \pi} \, G(\theta) .
\eeq
Trading $\int_{-\infty}^\infty d \theta$ with $\int_{\vep_0}^\infty d \vep$ one can show 
\beq
\label{cLr}
c_{UV} = \frac{6 s}{\pi^2}   \,  {\rm Lr}_2 ( s\, e^{-\vep_0} )
\eeq
where ${\rm Lr}_2$ is  Roger's dilogarithm:
\beq
\label{Rogers}
{\rm Lr}_2 (z) = {\rm Li}_2 (z) + \tfrac{1}{2} \log |z| \log(1-z)
\eeq
with ${\rm Li}_2$ the dilogarithm 
\footnote{Note $\Li_2 (1) = - 2 \Li_2 (-1) = \zeta(2) = \pi^2/6$, which is the origin of all factors of $6/ \pi^2$ throughout this article.}. 
\beq
\label{Li} 
{\rm Li}_2 (z) =  \sum_{n=1}^\infty \frac{z^n}{n^2} .
\eeq

For the sinh-Gordon model, $s=-1$ and $\htilde=1$.   The solution to \eqref{trans} is $\vep_0 = -\infty$, and this leads to $c_{UV}=1$ as expected.

\section{$T \Tbar$ perturbation of the sinh-Gordon model}

In this section we consider the action
\beq
\label{TTpert}
\CS = \int  d^2 x \( \inv{8 \pi} \d_\mu \phi \, \d_\mu \phi + \lambda \, \cosh (b \phi )  +\frac{\alpha}{\pi^2}  \,T \Tbar \) .
\eeq

\def\ghat{g}

\subsection{Qualitative renormalization group flow}

The dimensions of the couplings are $\dim{\lambda} = 2 + b^2$ and $\dim{\alpha}= -2$,  thus $\cosh (b\phi)$ is relevant and $T \Tbar $ is irrelevent.  
We can write 
\beq
\label{const}
\lambda = m^{2 + b^2}
\eeq
up to a multiplicative constant.  
Also 
\beq
\label{alphag}
 \alpha \propto  \ghat /m^2 ,
\eeq
again up to an overall constant, 
where $\ghat$ is the  dimensionless constant  in the CDD factor below \eqref{Scdd} and $m$ is a physical mass such as the mass of the sinh-Gordon particle.   
The constant of proportionality in \eqref{alphag} will be determined below for the massless case (see \eqref{params}).      

At very high energy $E\gg m$,  we can let $m\to 0$.   Here $\lambda \to 0$.   Thus in the   UV, the theory is dominated by 
\beq
\label{TT2}
 S = \int d^2 x \( \inv{8 \pi} \d_\mu \phi \, \d_\mu \phi  + \frac{\alpha}{\pi^2}  \, T \Tbar \) .
\eeq
On the other hand,  at very low energy,  $m\to \infty$ and $\alpha \to 0$, and the theory is dominated by the usual sinh-Gordon model.    
The latter is massive and as $m\to \infty$ at even lower energies the theory is empty with $c_{IR} =0$.   
Without the $T\Tbar$ perturbation  the sinh-Gordon model has $c_{UV}=1$.  On the other hand,  with the $T\Tbar$ it is as yet unclear if
the theory is UV complete in the sense that it has a local quantum field theory description in the UV.  This important issue will be discussed  in more detail below.  
In this article by ``UV complete" we mean that the UV behavior is controlled by a relevant perturbation of a CFT.  One should  point out that in the string theory context the perspective can be different. It is  believed that pure $T\Tbar$ deformations indeed have a  non-local stringy kind of UV completion,  where   the short distance physics is not governed by a local fixed point CFT.   (See the references in the Introduction.)    This is signified by a Hagedorn density of states. The UV completion is then interpreted as JT gravity in flat space minimally coupled to the seed quantum field theory \cite{Dubovsky2}. In this article we try to  limit our perspective to that of local quantum field theory.   

A qualitative description of the complete  
RG flow can thus be pictured as follows.   At very high energies the theory is well described by the $T \Tbar$ perturbation of
a free boson \eqref{TT2}.    Thus far $c_{UV}$ is unknown since it is not yet clear that \eqref{TT2} has an ultra-violet completion.    
Lowering the energy,  the model \eqref{TT2} consists of a massless particle and reaches an IR fixed point at some energy $E_1$ with $c=1$.    Lowering the energy 
further to much lower than $E_1$,  model \eqref{TTpert} is dominated by the sinh-Gordon model without the $T\Tbar$.   The latter has a UV fixed point with $c=1$.  At the lowest possible energies
there is an IR fixed point with $c_{IR} = 0$.    Thus the flow proceeds as $c(mR) = c_{UV} \to c(mR) \approx1 \to c_{IR}  =0$.   
The vicinity of $E_1$ where $c(mR) \approx1$   is a cross-over region where 
both perturbations in \eqref{TTpert} play a role and in fact compete.   Clearly there is no actual fixed point at $c=1$ unless one ignores one of the two perturbations.  
Nevertheless it is consistent to patch these flows together since $c_{UV}$  of sinh-Gordon and $c_{IR}$ of \eqref{TT2} are both equal to $1$.   
In other words, the model \eqref{TT2} flows to a free boson in its IR with $c=1$,  which then serves as the UV fixed point for the pure sinh-Gordon model.

\subsection{S-matrix and TBA}

Smirnov and Zamolodchikov have shown that the $T\Tbar$ perturbation of an integrable theory continues to be integrable.    Furthermore for the integrable case they showed that the  new S-matrix is the original S-matrix multiplied 
by a CDD factor.  
This CDD factor was originally proposed in \cite{DubovskyTBA}.   In our case this implies that the S-matrix for \eqref{TTpert} is 
\beq
\label{Smatrix2}
S(\theta) = S_\shG (\theta) \, S_\cdd (\theta) .
\eeq
The CDD factor must satisfy the usual constraints of unitarity $S_\cdd (\theta) S_\cdd(-\theta) =1$ and  crossing symmetry $S_\cdd (i \pi - \theta) = S_\cdd (\theta)$.    
It was proposed in \cite{SmirnovZ}  that $S_\cdd$ in rapidity space  has the simple form 
\beq
\label{Scdd}
S_\cdd (\theta) = \exp ({i g \sinh \theta}) .
\eeq
In the TBA framework,  the above expression for $S_\cdd$  defines the dimensionless constant $g$ which necessarily  vanishes when $\alpha =0$.  
Note that $S_\cdd (0) = 1$ so that the statistical parameter $s$ is unchanged by the perturbation.  Here, for the sinh-Gordon model $s=-1$.    

The TBA equations are the same as in Section IIIA with $s=-1$ and with the simple kernel 
\beq
\label{newG}
G(\theta) = G_\shG (\theta) + G_\cdd (\theta) 
\eeq
where 
\beq
\label{Gcdd}
G_\cdd (\theta) = g \cosh \theta .
\eeq
These TBA equations in principle describe the complete renormalization group flow including the cross-over region around $c(mR) \sim 1$ since it now depends on both of the  
dimensionless parameters $mR$ and $g$~\footnote{We hope to present a numerical study of this TBA equation elsewhere.  Here we are mainly interested in the pure $T\Tbar$ 
perturbation in the deep UV.}.

\section{The deep ultra-violet:  ~$T\Tbar$ perturbation of a free massless boson or fermion.}

Of particular interest is the model \eqref{TT2} since it describes the deep UV of the perturbed sinh-Gordon model.
The unperturbed theory ($\alpha =0$) is simply a free  massless boson.   This requires the statistical parameter $s=1$,  which is consistent with 
$S_\cdd (0) =1$, otherwise one would not obtain $c_{IR} =1$.  We are assuming that the $T\Tbar$ interactions are too soft to change the statistics to $s=-1$.  As for all irrelevant perturbations,  there is a fixed point  in the IR,  and this must be a theory of massless particles, otherwise $c_{IR} =0$ as
$m \to \infty$.  
We thus follow the description in Section IIIB for massless particles,  where L and R pseudo-energies were introduced.  

The free massless theory has $S_{LL} = S_{RR} = 1$,  which gives  $c_{IR} = 1$.      
For this reason it appears incorrect to  simply include $S_\cdd$ in $S_{LL}$ or $S_{RR}$ since this would  correspond to a conformal field theory,  which it is not the case when
$g\neq 0$. 
More importantly, with the latter choice  the massless TBA equations would not converge to a solution.    To see this,  define $L[f(\theta)] = \log \( 1- \exp ({-f(\theta)} )\)$ for
general $f(\theta)$.  
Suppose the kernel
is $G_\cdd$ given in \eqref{Gcdd}.  
For a massless TBA the convolutions in the integral equation involve 
$ g \int_{-\infty}^{\infty}  d \theta   \cosh (\theta)  L[e^\theta]$  to first order in a recursive solution,  which does not converge.   
On the other hand in a massive theory  as in the last section, we encounter 
$g \int_{-\infty}^{\infty} d \theta   \cosh (\theta)  L[\cosh \theta ]$ which does converge.

For the TBA system in this section and the next   we will need the convergent integral
\beq
\label{convInt}
 \int_{-\infty}^\infty d \theta \, e^\theta \, \log \( 1- sz\, e^{-y \,e^\theta} \) = - \frac{ \Li_2 (sz) }{y}, ~~~~ \Re (y) >0.
\eeq
We have included the fermionic case $s=-1$  since we will need it below;  for  most of the remainder of this section we deal only with bosons with $s=1$,  but will easily generalize it to fermions below.

The second option is to incorporate the CDD factor  into L-R scattering,  which is consistent with a breaking of conformal invariance.     In order for the TBA equations to converge to a solution,  it appears necessary to factorize the CDD factor:
\beq
\label{LR}
S_\cdd (\theta) = S_{LR} (\theta) S_{RL} (\theta),  ~~~~~~~~~~ 
S_{RL} (\theta ) =  \exp \( ig  e^\theta /2 \), ~~S_{LR}  (\theta) = \exp \( -i g e^{-\theta} /2 \) . 
\eeq
This second option is somewhat unexpected and suggests that perhaps a rethinking of massless scattering in this context is warranted.  
In this paper we simply show that this prescription leads to correct results obtained by other rigorous methods,  such as  based on the more general inviscid Burgers equation.

The kernels which follow from these S-matrices, $G(\theta) = -i \d_\theta \log S(\theta)$, are 
\beq
\label{GLR} 
G_{RL} (\theta) = G_{LR} (-\theta) =  g  \, e^\theta /2 .
\eeq
This leads to the now coupled TBA equations
\barray
\vep_R (\theta_R) &=&  \tfrac{mR}{2} e^{\theta_R} + \,G_{RL} \star \log \( 1- e^{-\vep_L (\theta_L)} \)  \cr
\vep_L (\theta_L) &=&  \tfrac{mR}{2} e^{- \theta_L} + \,  G_{LR} \star \log \( 1- e^{-\vep_R (\theta_R)} \)  
\label{TBAmassless2}
\earray
and $c = c_L + c_R$ has  the same expression  as before \eqref{cLR}.  Above it is implicit that the convolution with $G_{LR}$ involves $G_{LR} (\theta_L - \theta_R)$ 
and similarly with $G_{RL} (\theta_R - \theta_L )$.  
The convolution integrals in the above TBA equations now involves $g \int_{-\infty}^\infty d \theta \, e^{\theta} \, L[ e^\theta ] $ to lowest order which does converge. 
 When $g=0$ this TBA system is just that of a free massless  boson with $c=1$,  which implies
$c_{IR} = 1$.

The coupling between $\vep_L$ and $\vep_R$ via $G_{LR}$ and $G_{RL}$ breaks the conformal  invariance.    One can no longer remove the $mR$ dependence by
shifting $\theta_{L,R}$ since the kernels involve $\theta_L - \theta_R$.     As explained above,   $m$ is not  the mass of a physical particle,  however it is physically meaningful  in this massless theory since it is needed to set an energy scale. Unlike the massive case where $c$ depends on $mR$ and $g$ independently,  here   we expect $c$ to only depend on the combination
\beq
\label{hdef} 
h \equiv  \frac{g}{(mR)^2}  ~ \propto ~\alpha/R^2 
\eeq
since $\alpha$ is the only parameter of the theory.    The constant of proportionality will be fixed below.   

One can solve the above TBA equations exactly.    It is easy to show that
\beq
\label{epLR}
\vep_R (-\theta ) = \vep_L (\theta ), ~~~~~c_L = c_R.
\eeq   
One can then eliminate $\vep_L$ from the first equation in \eqref{TBAmassless2} to obtain
\beq
\label{vepRonly}
\vep_R (\theta) =  \tfrac{mR}{2}   e^{\theta}  +  \tfrac{g}{4\pi} \, e^\theta  \int_{-\infty}^\infty d\theta'  \,  e^{\theta'} \log \( 1- e^{-\vep_R (\theta')} \) .
\eeq
It is clear that the solution is of the form 
\beq
\label{vepRA}
\vep_R (\theta) =  
A\,\tfrac{mR}{2} \, e^\theta 
\eeq
for a constant $A$ that depends only on $mR$ and $g$.    Inserting the above form into \eqref{vepRonly},   $A$ must satisfy the quadratic equation 
\beq
\label{Aeq}
A = 1 - \frac{\pi g}{6(mR)^2} \inv{A} .
\eeq
There are two roots to this equation,  and we choose the one that gives the correct IR limit:
\beq
\label{Asol}
A = \inv{2} \( 1+ \sqrt{ 1-  \frac{2 \pi g} { 3 (mR)^2} } \) .
\eeq
Inserting this into \eqref{cLR} 
where $c=c_L + c_R = 2 c_R$ one obtains
\beq
\label{cTT}
c(h) = \frac{2 c_{IR}}{1 + \sqrt{ 1-   \frac{2 \pi h}{3}  c_{IR}}  }
\eeq
with $c_{IR} =1$.   

Repeating the above analysis for fermions ($s=-1$)  one obtains the same result as \eqref{cTT} but with $c_{IR} = 1/2$;  see Section VII. 
In fact in Section VII we introduce a continuously varying $c_{IR}$ into the TBA via a chemical potential and show that the result \eqref{cTT} continues to hold.  

We can now finally compare the above TBA result with  previously known results on the ground state energy $E (R)$ that were obtained by rather different methods,  in particular 
by utilizing the inviscid Burgers equation \cite{SmirnovZ,Tateo1}.    This comparison serves to verify our proposed TBA equations and also fix conventions for 
the couplings $g$ and $\alpha$.      Using the conventions in \cite{SmirnovZ}, the ground state energy satisfies the differential equation 
 \beq
 \label{Burgers}
 \d_\alpha E + E\,  \d_R E = 0 .
 \eeq
 Indeed, one finds that \eqref{cTT} satisfies the above differential equation if one identifies
 \beq
 \label{params}
 h = - \frac{\alpha}{R^2}, ~~~~~\Longrightarrow ~~ g = - \alpha \, m^2.
 \eeq

\section{Renormalization group flows}

In this section we attempt to interpret the RG flow for the models of the previous section, namely the $T\Tbar$ perturbation of the free boson or fermion.       
In the next section we will derive a more general  $c(h)$ for a model with a continuously varying $c_{IR}$ with the result \eqref{cTT},  
where $-\infty < c_{IR} < \infty$.    We thus base our discussion on the universal result \eqref{cTT}.  

\subsection{General observations} 

For conventional massless flows,  the ultra-violet and infra-red fixed points are known,  with known $c_{UV}$ and $c_{IR}$ \cite{ZZ,Fendley}.      A perturbation of the UV fixed point 
by a relevant operator leads to a renormalization group (RG) flow  to the IR fixed point,  where the flow arrives to the latter via an irrelevant operator.  Namely, in the deep 
IR  the theory is described by $\CS = \CS_{IR} + \gamma  \int d^2 x \, \CO (x) $, where the scaling dimension of the local field  $\CO (x) $ is $\geq 2$.  For instance,  for the $O(3)$ non-linear sigma model with topological
angle $\vtheta = \pi$, 
$c_{UV} =2$ and one reaches the $c_{IR} =1$  fixed point  via the marginally irrelevant operator $\CO  = \sum_{a=1}^3  J^a {\bar{J^a}}$,  where $J^a$ are $SU(2)$ currents.  Throughout the flow the $SU(2)$ symmetry is preserved.    If one is only given an effective description of the theory near the IR,  in general one cannot reconstruct the reversed flow to the UV based soley  on this information.   
The possibility remains that a reversed flow to a well defined UV theory  does not exist and 
the theory is said to be asymptotically incomplete.    In  4 dimensions, this issue is central to the difficulties encountered in quantum gravity for instance. 
 
An important aspect  to consider for  RG  flows is Zamolodchikov's c-theorem \cite{ctheorem}.   
Generally,  given a theory with generic coupling $h$,  one has
\beq
\label{dc}
\frac{dc}{d \log r} = \beta (h) \frac{dc}{dh}, ~~~~~~r \equiv mR
\eeq
where $\beta(h) =  dh/ \log r$ is the RG beta function.   Recall that the IR (UV) limit corresponds to $r\to \infty $ ($r \to 0$).  
Zamolodchikov's construction of a c-function assumes that in the UV the  theory is a relevant perturbation of a well defined CFT.   
This is not necessarily valid in the present context.  
 However our definition of a $c$-function is the most natural in the context of the TBA.  
It would seem unlikely that one can prove a c-theorem if in the UV the theory is not a relevant perturbation of a CFT.
Nevertheless it is instructive to consider the implications of the c-theorem if it were valid.   
 The c-theorem states that $c(r)$ decreases in the flow to the IR: ~ $dc/d\log r \leq 0$.   
 Fixed points where $dc/d \log r =0$  can actually arise in two ways:   either $\beta (h) =0$ or $dc/ dh =0$, where the former is more typical.   
 As we now explain,  the massless flow of $c(r)$ does not appear to  have a UV completion in all cases.      

For our model,  $\beta (h) = -2 h$ thus 
\beq
\label{dc2}
\frac{dc}{d \log r} = - \frac{4 \pi h}{3} \,   \frac{ (c_{IR})^2}
{\( 1+\sqrt{1 - \tfrac{2\pi h}{3} c_{IR} } \)^2 \sqrt{1 - \tfrac{2\pi h}{3} c_{IR}} } .
\eeq
Recall that asymptotic RG flows correspond to 
\beq
\label{RGcases}
{\rm RG ~ flow}~~
\begin{cases}
{\rm toward ~ the~ IR:} ~~~~ r\to \infty  ~~\Longrightarrow~~ h \to 0 \\
{\rm toward ~ the~ UV:} ~~~ r\to 0 ~~~ \, \Longrightarrow~~ h \to \pm \infty
\end{cases}
\eeq
If we require that the c-theorem is not violated,  then this requires $g\geq 0$ such that $h\geq 0$.     Let us assume this for the remainder of this subsection.  
Thus $h=0$ is a line of IR fixed points where $\beta (h) =0$ and $-\infty < c_{IR} < \infty$ is identified as the value of $c(h=0)$.  
If $c_{IR} < 0$ then $h$ can be taken to infinity.   
Thus there is only one class of flows with continuous fixed points in both the IR and UV that is also consistent with the c-theorem:   

\bigskip
\noindent
{\bf Flows with continuous IR and UV fixed points:}   At the IR fixed point,  $h=0$ and   $c_{IR} < 0$.  The flow is to   $c_{UV} =0$  at $h = h_* =  \infty$ regardless of the value $c_{IR}$.    
In the limit $h \to \infty$,  $c \sim 1/\sqrt{h}$.  
If $c_{IR} =0$ then $c_{UV} =0$ and the theory is interpreted as being conformally invariant since $c$ does not flow at all,  i.e. it is independent of $h$.    
An example of such a flow  with $c_{IR} = -1$ is shown in Figure 1.
\bigskip

\begin{figure}[t]
 \label{cNeg}
\centering\includegraphics[width=.6\textwidth]{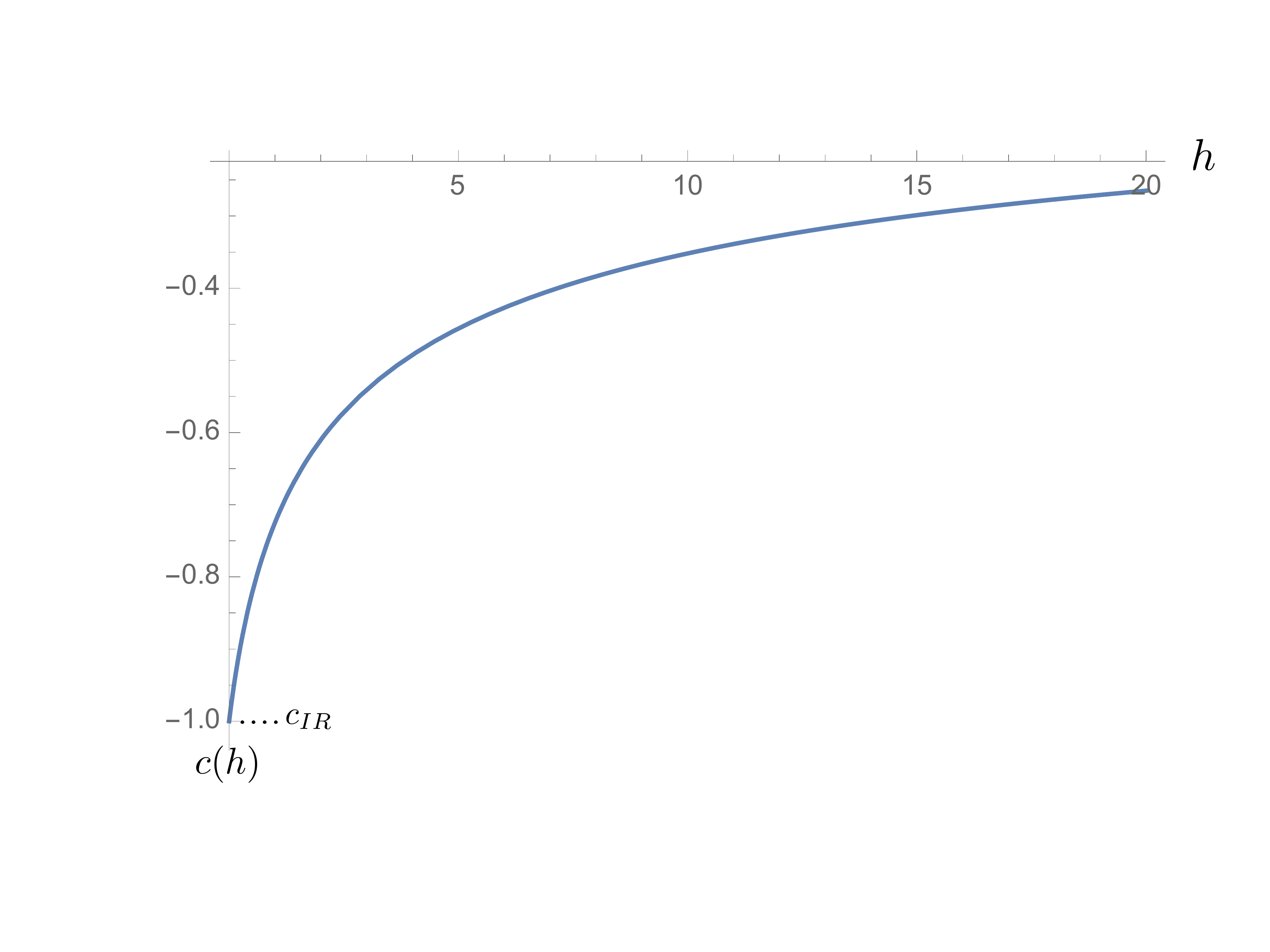}
\caption{$c(h)$ given in \eqref{cTT} with $c_{IR} =-1$ which corresponds to fugacity $z=-2.393..$ of the next section.  
 }
\end{figure}

\def\cstarUV{c^*_{UV}}

It remains to interpret the physically more interesting  $h>0$ flows when $c_{IR} > 0$, which is more delicate and requires a proper interpretation.     Here $r$ cannot be arbitrarily small:    if $r < r_*$ then $c(h)$ develops an imaginary part.   Thus this flow only
makes sense if 
\beq
\label{rstar}  
r < r_*, ~~{\rm where} ~~ \frac{g}{r_*^2} = h_* \equiv \frac{3}{2 \pi c_{IR}} ,
\eeq
where this equation defines $mR_* = r_*$ and $h_*$.  
This suggests a theory with a shortest possible distance,  such as a discrete space-time, a string theory,  or a lattice approximation to a continuous theory.   
Clearly $h=h_*$ is not a continuous fixed point, and the theory at $h_*$ is not a CFT.     In fact $dc/ d\log r = -\infty$ there.   
In short,  one should conclude that in this case  the theory is not UV  complete in the usual sense of a local quantum field theory.  
Presumably in the string theory context  $r_*$  is related to the non-locality scale  beyond which local field theory techniques break down.
 Similar features have been observed in \cite{Kutasov} for  the case of single trace $T\Tbar$ deformations while analyzing the holographic c-function. There in their analysis the non-locality scale is proportional to the string length.
The  c-function in \eqref{cTT} developing an imaginary part at short distances  has also been observed in 
 \cite{Chakraborty} in analyzing the thermodynamic free energy.  Presumably this signifies the same physics since this square-root singularity is rather generic in this context. 
 See also \cite{MussardoSimon}.

      Nevertheless $c(h)$ is finite at $h_*$ and 
  we will refer to $c(h_*)$ as $\cstarUV$.       The next subsections will provide some  insights into the nature of the theory at $\cstarUV$ if it indeed exists.   Let us summarize:

\bigskip

\def\bdual{\tilde{b}}

\noindent
{\bf Flows with a discontinuity:}   We  consider  $h>0$ such that the c-theorem is not violated.  Here $c_{IR} >0$ and the flow terminates at $r_*$ defined in \eqref{rstar}.     As $h \to h_*$,  $c \to \cstarUV$ and 
$dc / d \log r \to \infty$,  however   $\cstarUV$ remains finite:
\beq
\label{cIRUV}
\cstarUV =  2 \, c_{IR} .   
\eeq
If  $c_{IR} =0$, then $\cstarUV =0$ in accordance with the continuous fixed point flows described above with $c_{IR} <0$.   Such a flow is shown in Figure 2 for
$c_{IR} =1$;  flows for other $c_{IR} > 0$ are very similar but with different $h_*$ and $\cstarUV$.

\bigskip

It is somewhat interesting, but just a bit,  to express \eqref{cIRUV} in terms of the Liouville coupling $b$  using \eqref{cbc}, since the general equation \eqref{cTT} is loosely motivated by the Liouville theory in the next section.     For simplicity we consider only the region $c_{IR} \geq 25$ where
$b$ is real.   Letting $b_{IR}$ and $b_{UV}$ be the IR and UV values of $b$, one finds
\beq
\label{bUV}
b_{UV} =
\begin{cases}
1/\sqrt{3} ~~~~~~~ \,\, {\rm as} ~~c_{IR} \to 25 \\
b_{IR}  / \sqrt{2}~~~~~{\rm as} ~~c_{IR} \to \infty, ~~{\rm with} ~~ b_{UV, IR} \to 0 \\
\sqrt{2}\,\, b_{IR} ~~~~~ \,{\rm as} ~~ c_{IR} \to \infty, ~~{\rm with}  ~~b_{UV, IR} \to \infty .
\end{cases}
\eeq
The second two relations are related by duality $b\to \bdual = 2/b$.   Performing a duality transformation on $b_{IR}$ only, one obtains 
\beq
\label{bIRdual}
b_{UV} =\inv{ \sqrt{2}} \,\, \bdual_{IR} = \frac{\sqrt{2} }{b_{IR}}~~~~~~~{\rm as} ~~c_{IR} \to \infty ~~{\rm with} ~~ b_{IR} \to \infty, ~ b_{UV} \to 0 .
\eeq
This should be compared with $b_{UV} = \bdual_{IR}$ which would incorrectly imply $\cstarUV = c_{IR}$.

\begin{figure}[t]
 \label{cOne}
\centering\includegraphics[width=.8\textwidth]{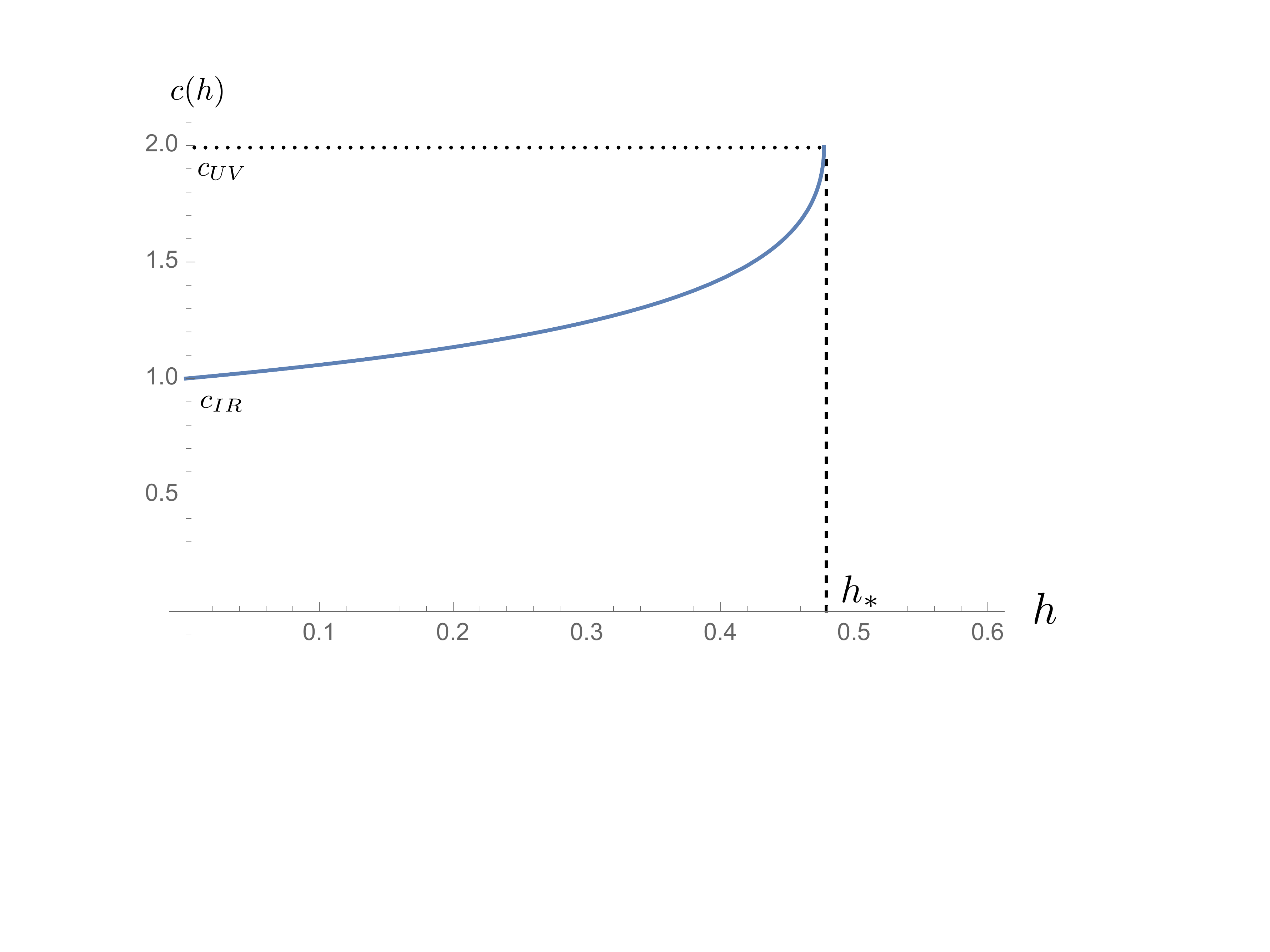}
\caption{$c(h)$ given in \eqref{cTT} with $c_{IR} =1$,  which is the model in Section VI.   In this figure $\cstarUV$ is labeled simply as $c_{UV}$.  This corresponds to fugacity $z=1$  of the next section.  
 }
\end{figure}

\def\meff{m_{\rm eff}}

\subsection{Discontinuity  as a  tachyonic vacuum instability}

Let us refer to the  above discontinuity  simply as a square root singularity.  
Based on  the form of the pseudo-energy $\vep$ in \eqref{vepRA} one can interpret $A \,m$ as an effective mass $\meff$:
\beq
\label{meffdef}
\meff = A \, m .  
\eeq
One sees that for $h> h_*$, $\meff$ acquires an imaginary part.   In the deep UV, $\meff$ becomes imaginary
\beq
\label{tach}
\lim_{h \to \infty}  \meff = i \, \sqrt{\frac{\pi h c_{IR}}{6} }\,\,m .
\eeq
A fictitious particle with an imaginary mass is commonly referred to as a tachyon,  however physically  it can signify a vacuum instability. For instance in the Higgs sector of the Standard Model the
Higgs potential has an imaginary mass, and one must perturb around the true vacuum which leads to a real physical  mass.   This tachyonic instability leads to square-root singularities since a single tachyon energy  of mass $m = i |m|$ becomes 
$E=\sqrt{\vec{p}^2 + m^2} = \sqrt{ \vec{p}^2 - |m|^2}$.    A similar kind of singularity was found for different models in \cite{MussardoSimon};  
in fact one can argue that this was the first work where such singularities were observed,  although the motivations were different.    

Further support for this interpretation of the singularity is as follows.   The parameter $h$ is proportional to $g$ which is proportional to $\alpha \, m^2$,  as in \eqref{alphag}.  
Thus making $m$ tachyonic,  $m \to i m$,  flips the sign of $h$.    It turns out that the flows with negative $h$ and $c_{IR} >0$ are resolved without singularities,  i.e.
 they are continuous and complete since they can be extended to all scales, 
with $c_{UV} = 0$.    See Figure 3.   It is important to reiterate that these flows for negative $h$ violate the c-theorem since $c_{IR} > c_{UV}$,  however they are still interesting.


\begin{figure}[t]
 \label{hNegative}
\centering\includegraphics[width=.7\textwidth]{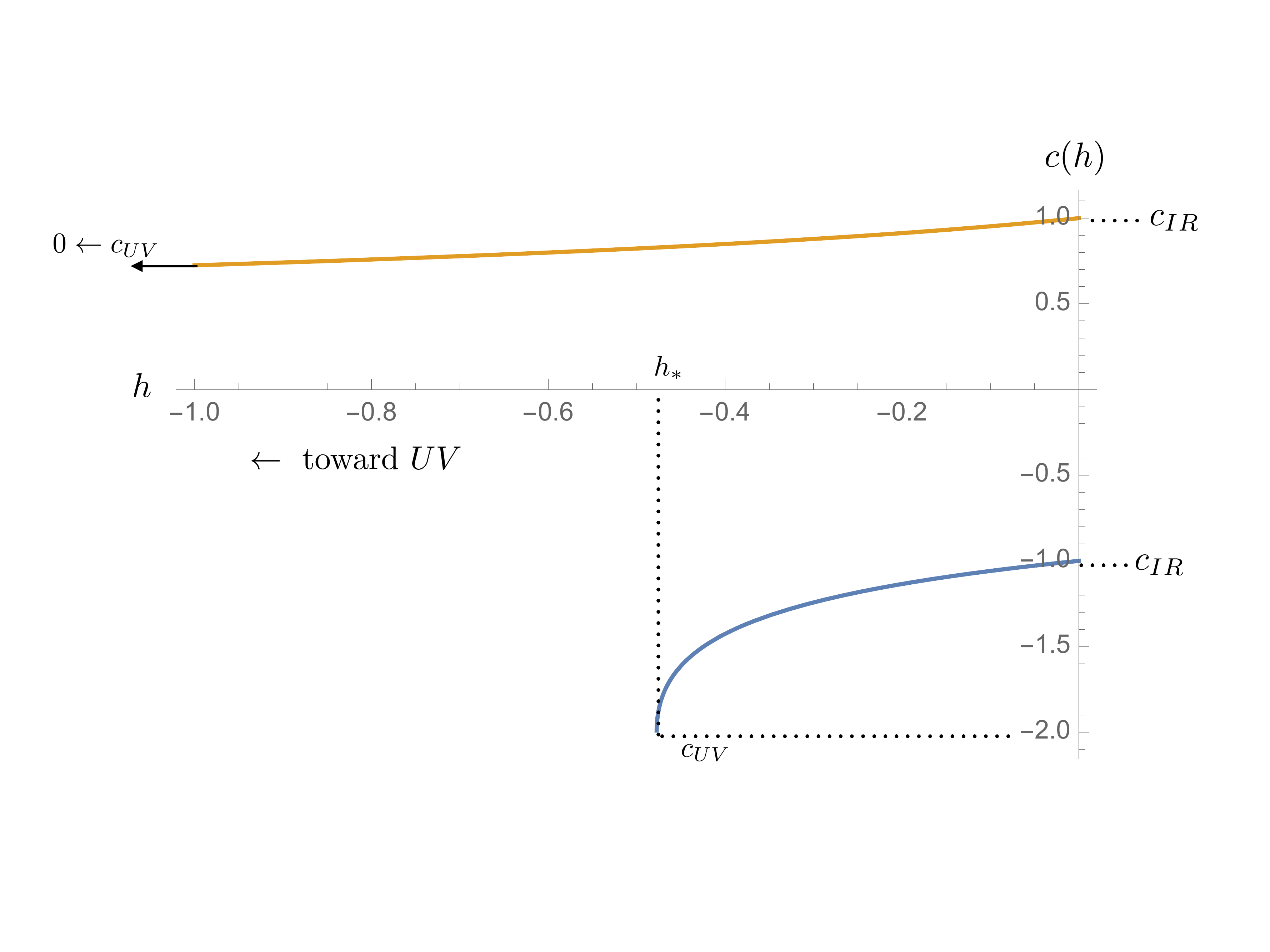}
\caption{Flows toward the UV for negative $h$ and $c_{IR}$ positive verses negative.  $c(h)$ is  given in \eqref{cTT}, and $c_{UV}$ equals $\cstarUV$ described in the text.    
 }
\end{figure}

\def\fp{{\rm fp}}

\subsection{Scaling dimension of the perturbing operators near the fixed points and the discontinuity}

\def\Sfp{S_\fp}
\def\Opert{\CO_{\rm pert}}

From the behavior of $c(r)$ near an IR or a UV fixed point CFT,  we can determine the scaling dimension of the operator which perturbs the RG flow away or toward the fixed point,
and this will provide important information on the nature of the UV fixed point and also the singularity at $h_*$.      
Near a fixed point (fp) the theory is described by the action
\beq
\label{Sfp}
S= \Sfp + \gamma \int d^2 x ~ \Opert (x) 
\eeq
where $\gamma$ is a coupling.   The scaling dimensions satisfy
\beq
\label{scaling} 
\dim{\gamma} = 2 - \dim{\Opert} .
\eeq

Let us use the language of statistical mechanics based on the free energy density $\CF$ at temperature $T$  in \eqref{cR}  with  $R\equiv 1/T$.  One has
\beq
\label{FreeZ}
\CF = - \frac{T}{V} \log Z
\eeq
where $V$ is the one dimensional volume which is the length of the cylinder,  and $R$ its circumference.   Formally $Z$ is a partition function
\beq
\label{Zpart}
Z = \int  D \Phi ~ e^{-S[\Phi]} 
\eeq
where $\Phi$ are the fields that define the theory.  
To lowest order in $\gamma$,  
\beq
\label{Zpert}
Z \approx Z_\fp \cdot \exp \( - \gamma \int d^2 x \,\, \langle \Opert \rangle_T \)
\eeq
where  $\langle \Opert \rangle_T$ is a finite temperature 1-point function.   The latter can be computed using form factors \cite{LecMuss},  however we won't need 
such a formula here.   
From the definition \eqref{cR}, to lowest order in $\gamma$ 
\beq
\label{deltac}
c(r) = c_\fp + \delta c, ~~~~~~\delta c = -\gamma \( \frac{6}{\pi} \frac{R}{V} \int d^2 x  \,\,\langle \Opert  (x) \rangle_T \),
\eeq
where $c_\fp$ is the constant value of $c(mR)$   at the fixed point.  
It follows from \eqref{scaling} that $\gamma \propto m^{2-\dim{\Opert}}$.   Since $c(r)$ is dimensionless,  one must have
\beq
\label{delc}
\delta c \propto r^{2 - \dim{\Opert}} .
\eeq

Using the above formula \eqref{delc} we proceed to analyze the two kinds of fixed points,  and also the discontinuity.     

\bigskip\bigskip

\noindent
{\bf IR fixed points.} ~~~ Here we expand $c(h)$ in \eqref{cTT} around $h=0$:
\beq
\label{cIRpert}
c(h) = c_{IR} +  h\,\(  \frac{\pi c_{IR}^2}{6} \) + h^2\, \( \frac{\pi^2 c_{IR}^3}{18}\) + O(h^3)
\eeq
Since $h=g/r^2$, $\delta c \propto r^{-2}$ thus we verify that  $\dim{\Opert} = 4$,  which is the dimension of $T\Tbar$.   

\bigskip
\noindent
{\bf Continuous UV fixed points.}~~~
Here $c_{IR} \leq 0$, $h \to h_*= \infty$  and $c_\fp=c_{UV}=0$.   Expanding around $h=\infty$:
\beq
\label{cUVInfinity}
c(r) =  - h^{-1/2} \sqrt{|c_{IR}|} +  \frac{3}{\pi h} + O(h^{-3/2} ). 
\eeq
Thus $\delta c \propto r$ which implies 
\beq
\label{cUVone}
\dim{\Opert} = 1
\eeq
for all values of $c_{IR}$.  
The fact that $\Opert$ is a relevant operator explains why both the IR and UV fixed points are continuous for the flow from $c_{UV}=0$  to $c_{IR} < 0$.

\bigskip
\bigskip

\noindent
{\bf Discontinuous flows.}~~~
Although the theory at $h_*$ is not a CFT fixed point,  one can still analyze its vicinity since $\cstarUV$ is finite.  
Here $c_{IR}>0$ and $\cstarUV = 2 c_{IR}$.   As discussed above,  the flow toward the UV  terminates at $h=h_* = 3/(2 \pi c_{IR})$, namely the flow cannot be extended to arbitrarily small $r$ 
with $h>h_*$.   However $c(h)$ at $h=h_*$ is well defined and finite,  thus we can expand $c(h)$ around $h_*$ from below.  The result is 
\beq
\label{cUVstar}
c = 2 c_{IR} - \sqrt{\epsilon} \( \sqrt{ \frac{8 \pi}{3}} c_{IR}^{3/2} \) + \epsilon \( \frac{4 \pi c_{IR}^2}{3} \) + O(\epsilon^{3/2}), ~~~~~~\epsilon \equiv h_* -h .
\eeq
Thus $\delta c \propto \sqrt{\epsilon} \propto r^{-1}$.   This gives 
\beq
\label{Opert3}
\dim{\Opert} = 3,
\eeq
again for all values of $c_{IR}$.  

The above result  \eqref{Opert3} is interesting for several reasons.   The operator is more relevant than the $T\Tbar$ perturbation in the IR,  which perhaps can be 
anticipated.   However $\Opert$ is still irrelevant, whereas for a continuous fixed point in the UV  it would be
 relevant.   This suggests at least two interpretations.   Either the model has a
true physical shortest possible distance $r_*$ from which the flow toward the IR begins, and the operator that directs the flow is $\Opert$ with dimension $3$.  
   Or,  our analysis only provides a partial UV completion,  and the flow to the  deep  UV is incomplete  in  our description.   One possibility is a vacuum instability as described in the previous sub-section.  
In summary,  we are unable to fully resolve these  possibilities in this article 
[36]. 
%


\section{Varying the infra-red central charge $c_{IR}$}

\def\Li{{\rm Li}}

The goal of this section is to generalize the above results to a continuously varying $c_{IR}$ and to show that its $T\Tbar$ deformation leads to \eqref{cTT}.  
This will be accomplished below by introducing a chemical potential.   

Let us first motivate our construction based on the Liouville theory.    We emphasize that based on the simple considerations of this section we only establish a loose connection to Liouville and certainly do not even conjecture that our proposal really describes the deformed Liouville theory;   rather at this stage it should just be viewed as a toy model.
That being said, 
 consider the $T\Tbar$ perturbation of the Liouville theory defined by the 
 action
\beq
\label{LiouS2}
S = \int d^2 x \( \inv{8\pi} \d_\mu \phi  \, \d_\mu \phi + \hat{\lambda}  \,  \exp ({b  \phi} )  + \frac{\alpha}{\pi^2}  \, T\Tbar \) 
\eeq
where $\phi$ has a background charge such that $e^{b \phi}$ is an exactly  marginal operator when $\alpha =0$.   
Our viewpoint is to incorporate most of the properties one expects of such a model,  which already significantly constrains the possibilities.    
The most uncertain is  how to build in the background charge into the thermodynamic scattering theory.  
As previously stated we  cannot derive all of the features of the following proposal from
first principles based on the Liouville action.  If it turns out that the TBA proposed below has little to do with Liouville,  then our  construction is still a correct way to 
introduce a variable $c_{IR}$.  

\subsection{S-matrices}

Since the sinh-Gordon theory can be viewed as a perturbation of Liouville by an additional exponential operator $e^{- b \phi}$,  we assume the spectrum consists of a single scalar particle as for sinh-Gordon.    
The Liouville theory with appropriate background charge is a conformal field theory and $T\Tbar$ is an irrelevant operator, thus this particle should be massless.   
The first task is to find massless S-matrices that describe some properties of  the CFT of the Liouville theory before turning on $T\Tbar$, in particular the central charge $c$.  
One possibility is  to consider $S_{LL} = S_{RR}$ equal to the sinh-Gordon S-matrix, as was done for the O(3) sigma model in \cite{ZZ}.   The problem with this idea is that for the conformal version with  $S_{LR} =0$,  as shown in Section IIIC,  $c_{UV} = c_{IR} =1$ regardless of the
coupling $b$.    Thus it does not seem possible to incorporate a background charge leading to \eqref{cbc}  for all $b$ by an appropriate choice of $S_{LL}$ and $ S_{RR}$  for a single particle. This doesn't leave many options for modifying the TBA without changing the spectrum,  however we will propose one in the next sub-section based on a chemical potential.  

Long ago it was shown that the S-matrix for the Liouville theory is trivial at tree level,  i.e. $S=\pm 1$, such that the statistical parameter above is $s=S=\pm 1$ \cite{DHoker}.   This result was extended to the full quantum theory in 
\cite{Yoneya}.  
We take these results  on face value and set $S_{LL} = S_{RR} = 1$ or $-1$,  such that there are no $G_{LL}$ nor $G_{RR}$ terms in the  bosonic or fermionic TBA depending on whether $S = \pm 1$. 
Turning on the $T\Tbar$ perturbation, as in the previous section we identify $S_{LR}$ with the CDD factor  $S_{\cdd}$ in eqn. \eqref{Scdd} 
where $g =- m^2 \alpha$ with $m$ an energy scale.
 It remains to incorporate the background charge which is independent of the  CDD parameter $g$.   
 
 One should point out that
 a trivial S-matrix does not imply that the correlation functions are also trivial.  
  For instance, in the free Majorana fermion description of the Ising model,  the S-matrix for the fundamental fermions is $-1$,  however correlation functions of the spin field are complicated  due to its non-trivial N-particle form factors for all N. They satisfy non-linear Painlev\'e differential equations for instance.  The same occurs for fields $e^{i \alpha \phi}$ in the doubled Majorana, or Dirac,  version \cite{BL}.   For the Liouville theory much is known about such non-trivial correlation functions;  see the review  \cite{Teschner} and references   therein.

\def\cbose{c_{\rm bose}}
\def\cfermi{c_{\rm fermi}}

\subsection{Background charge as a chemical potential}

Consider first a free massless boson with no perturbations with $s=1$.   Let us include a chemical potential  $ \mu$ for both left and right movers.   
Specifically 
\beq
\label{mus}
\vep_R (\theta) =  \tfrac{mR}{2}  e^{\theta}  -\mu R ,~~~~~~~ 
\vep_L (\theta) = \tfrac{mR}{2} e^{-\theta }  -  \mu^* R, 
\eeq
and define the fugacity 
\beq
\label{fugacity}\
z \equiv e^{\mu R} .
\eeq
The integrals in \eqref{cLR} can be expressed in terms of the dilogarithm:
\beq
\label{cvtheta}
c (z) = c_L + c_R = \frac{3}{\pi^2} \( \Li_2 (z )  + \Li_2 (z^*) \)
\eeq
which is real. 
Although there are interesting cases with imaginary chemical potential,  fortunately it will be sufficient for us to only consider $z$ real such that
\beq
\label{cbose}
c (z) = \cbose(z)  \equiv  \frac{6}{\pi^2} \,  \Li_2 (z ) .
\eeq
\footnote{
For instance consider a pure phase $z=e^{i\vtheta}$.   There is a remarkable identity    
$$\cbose= 1- \frac{ 3\vtheta}{\pi} \( 1- \frac{\vtheta}{2\pi} \) $$
 valid for $ 0< \vtheta <2 \pi$.  Such a formula explicitly relates $c$ to $\vtheta$.  Due to the periodicity in $\vtheta$,  this choice of $z$ does not suit
our purposes since it is restricted to $-1/2 <c < 1$ which is not compatible with the duality $b\to 2/b$.  It would also be hard to justify $c_L  = c_R^* \neq c_R$.}.

The function $\Li_2 (z)$ has a branch cut in the complex $z$ plane that runs from $z=1$ to $z=\infty$ along the real axis.   We therefore restrict to $z\leq 1$:
\beq
\label{cbose2}
z \leq 1 ~~~~~ ~~  \Longrightarrow ~~~ -\infty < \cbose \leq 1
\eeq
The region $\cbose < 0$  corresponds to  $z< 0$,  which requires the  imaginary part of $\mu R$ to equal  $\pi$, however $z$ is still real.

In order to cover the full range of $c$,  interestingly  the particles must change from bosonic to fermionic statistics for $z\geq 1$.   
This could be anticipated since $z>1$ corresponds to a positive chemical potential $\mu$ which is usually associated to the existence of a Fermi surface;  in fact
at zero temperature, $\mu$ equals the Fermi energy.   
Let us incorporate the fact that in the field theory at zero temperature, a Dirac fermion with fields $\psi_\pm$ and $\bar{\psi}_{\pm}$ where
$\pm$ is the U(1) charge can be ``bosonized" with a single free boson where $\psi_\pm = e^{ \pm  i \vphi}$.    For zero chemical potential, $z=1$ and both the
bosonic and fermionic descriptions have $c=1$.    This leads us to double $c$ with the identification 
\beq
\label{cfermi1}
{\rm for} ~ z \geq 1  :~~~~~c(z) = \cfermi (z) \equiv - \frac{12}{\pi^2} \, \Li_2 (-z) .
\eeq
Since $\Li (-z)$ has no branch cut for $z>1$,  this leads to the range $1\leq \cfermi < + \infty$.  

In summary,  in order to cover the full range of $c$ one needs to distinguish  two cases: 
\beq
\label{ccases}
c(z) =
 \begin{cases}
  \cbose (z) ~~~{\rm if} ~ z\leq 1,~~~~~ \Longrightarrow~~~ -\infty <  \cbose \leq 1 \\
  \cfermi (z) ~~{\rm if} ~ z \geq 1,~~~~~\,  \Longrightarrow~~~~~~~~ 1 \leq  \cfermi  <  \infty .
\end{cases}
\eeq
With this definition,  these two regions can be joined since $c(z)$ is continuous at $z =1$ where the statistics changes.   However the derivative of $c(z)$ is discontinuous at $z=1$,  possibly signifying some kind
of phase transition.     Interestingly,  in this thermodynamic approach the usual field theoretic ``forbidden" zone $1\leq c \leq 25$ is just a smooth continuation of the  region 
with $c\geq 25$.   The value $c=25$ occurs at $z= 469.299...$ where $c(z)$ is smooth.      

The asymptotic behavior of $\Li_2 (z)$ gives approximate expressions for the inverse function:
\beq
\label{asym}
z(c) \approx 
 \begin{cases}
  e^{ \pi \sqrt{(c-2)/6} } ~~~~~~~{\rm as} ~~ c\to \infty \\
 - e^{  \pi \sqrt{(-c-1)/3} }  ~~~{\rm as} ~~ c\to -\infty .
\end{cases}
\eeq
For $c=25$ this gives $z\approx 469.136$,  indicating that $c=25$ is well into the asymptotic regime.  
There is no simple  expression  for the Liouville coupling $b$ as a function of $z$ in general, especially in the  region  $1\leq c \leq 25$ where $b$ is complex.  
However let us point out that, as discussed above,  all the continuous fixed points with $c_{IR} < 0$ have $c_{UV} =0$, and  by \eqref{aa02} the latter corresponds
to $\beta =2/\sqrt{3}$. The latter  turns out to be the $\CN =2$ supersymmetric point of the sine-Gordon model (before turning on a background charge) \cite{Fractional}.  
  
\subsection{TBA equations for the $T\Tbar$ perturbation and their exact solution}

Now turn on the $T\Tbar$ perturbation so that $h\neq 0$.   
For this  irrelevant perturbation the central charge $c(z)$ in the last sub-section is identified with the IR 
value $c_{IR}$.   Thus,  below,   $z$ is implicitly  chosen as the  function $z(c_{IR})$ where $z(c)$ is the inverse of the function $c(z)$
\beq
\label{zofc}
z(c(z')) = z'
\eeq 
where  $c(z)$  is given in \eqref{ccases}.  Thus  $z(c)$ depends on whether $c$ is in the bosonic verses fermionic regime. 
    As in the last section the $T\Tbar$ perturbation introduces an energy scale which $c$ is a function of,  and the 
TBA can be used to study this dependence.    

First consider the bosonic case where $z\leq 1$.   
With the above choice of S-matrices, 
the TBA equations are as in \eqref{TBAmassless2} but with the 
sources $mR \, e^{\pm \theta}/2 \to mR \, e^{\pm \theta}/2 -\mu R$.  
Shifting $\vep_R$ and $\vep_L$  by $-\mu R$  one obtains 
\barray
\vep_R (\theta_R) &=&  \tfrac{mR}{2} e^{\theta_R} + \,G_{RL} \star \log \( 1- z e^{-\vep_L (\theta_L)} \)  \cr
\vep_L (\theta_L) &=&  \tfrac{mR}{2} e^{- \theta_L} + \,  G_{LR} \star \log \( 1- ze^{-\vep_R (\theta_R)} \) 
\label{TBAmassless3}
\earray
where $z=e^{\mu R}$ 
and the kernels are the same as \eqref{GLR}.  
The formulas for $c_{L,R}$ in \eqref{cLR} still apply with $s=1$ but with $z$ dependence:
\beq	
\label{cRz}
c_R   = - \frac{3 mR}{2 \pi^2}  \int_{-\infty}^\infty d \theta \, e^\theta \log \( 1-z e^{-\vep_R} \)
\eeq
and similarly for $c_L$.  
It is straightforward to show that 
$\vep_L (\theta) = \vep_R (-\theta)$,  such that $c_L = c_R$.

\def\hhat{\hat{h}}

We proceed to exactly solve the TBA equations  as in the previous section.  
Eliminating $\vep_L$ from the above coupled equations and using the properties of the kernel \eqref{GLR} one finds
\beq
\label{vepRtba}
\vep_R (\theta) =  \tfrac{mR}{2} e^{\theta}  +  \tfrac{g}{4\pi} \, e^\theta \int_{-\infty}^\infty d \theta' e^{\theta'} \, 
\log \( 1 -z e^{-\vep_R (\theta')}    \) .
\eeq
As before the solution is of the form
\beq
\label{vepRA}
\vep_R (\theta) = A \, \tfrac{mR}{2}  \, e^\theta
\eeq
where $A$ is a constant that depends on $h= g/(mR)^2$ and $z$.  
 $A$ now satisfies the quadratic algebraic equation
\beq
\label{Aeq}
A = 1 - \frac{h}{\pi} \,  \frac{\Li_2 (z)}{A}  ,
\eeq
and one obtains
\beq
\label{Abranch}
A(h, z) = \inv{2} \( 1 + \sqrt{1 - \frac{4 h}{\pi} \, \Li_2 (z)} \) .
\eeq
Finally
\beq
\label{cLRz}
\cbose(h,z) = c_R + c_L  = \frac{6}{\pi^2} \(  \frac{\Li_2 (z) } {A(h,z)} \) .
\eeq
The latter can be entirely expressed in terms of $c_{IR}$ using $c_{IR} = 6\,  \Li_2 (z)/\pi^2$:
\beq
\label{cboseIR}
\cbose (h,c_{IR}) = \frac{2 c_{IR}}{1 + \sqrt{1-\frac{2 \pi h}{3} c_{IR}}}
\eeq

Similar results apply to the fermionic $z\geq 1$ case.   Recall the fermionic case is a doubled Dirac version,   with pseudo-energies 
$\vep_{L,R}^\pm$ where $\pm$ refers to the U(1) charge.   
There is some freedom in the definition of $h$ for the bosonic verses fermionic regimes,  since $m$ is not the mass of a physical particle,  and its relation to 
$\alpha$ is only known up to an overall constant.    For $z=1$,  the bosonic and fermionic $c(z)$ agree for $h=0$.   We require that they also agree at $z=1$ for arbitrary $h$. 
This requires  the rescaling $g \to 2 g$ in going from the bosonic to fermionic case,  so that $h\to 2 h$  and $s\to -1$ in the above formulas,  with $h$ still equal to $g/(mR)^2$.    The form of the 
TBA equations again  implies $\vep_R^\pm (\theta) = A \mr e^\theta $,  where now $A$ satisfies 
\beq
\label{Afermi1}
A = 1+ \frac{2h}\pi \( \frac{\Li_2 (-z)}{A} \)
\eeq
with solution 
\beq
\label{Afermi2} 
A (h,z)= \inv{2} \( 1 + \sqrt{ 1 + \frac{8h}{\pi} \Li_2 (-z) } \).
\eeq
The resulting central charge is 
\beq
\label{cferniIR}
\cfermi (h, c_{IR})  = - \frac{12}{\pi^2} \( \frac{\Li_2 (-z)}{A(h,z)} \) =  
 \frac{2 c_{IR}}{1 + \sqrt{1-\frac{ 2 \pi h}{3} c_{IR}}}
 \eeq
 where we have used 
 $c_{IR} = -12 \, \Li_2 (-z) /\pi^2$.   By design $\cbose (h,c_{IR}) = \cfermi (h, c_{IR})$ only where $c_{IR} =1$, since the latter is the unique point $z=1$
 where both the bosonic and fermionic values of $c_{IR}$  agree.

We  interpret the RG flow as discussed in detail  in the previous section.

\section{Conclusions}

In summary, 
we proposed Thermodynamic Bethe Ansatz equations for $T\Tbar$ perturbations of 
the massive sinh-Gordon model,  the massless free boson and fermion.    In the last section we showed how to incorporate a continuously variable $c_{IR}$.
  For the massless theories,  we  solved the TBA equations exactly and showed how they reproduce the previously known results for the ground state energy $E(R)$.  
  The novelty of our proposed TBA equations was a factorization of the CDD factor into Left and Right moving parts.        
Based on these exact solutions for the scale dependent central charge,  we studied the  renormalization group flows  towards the UV.     For the case of $c_{IR}>0$ and $\alpha < 0$,  which respects the c-theorem,  the flow has a singular point at $mR_*$.     For $c_{IR} < 0$ the flow extends all the way down to $R=0$ where $c_{UV} = 0$.    

Let us identify three  aspects of our analysis that deserve further study, scrutiny, and possible cross-checks.    
First,  the type of massless TBA we proposed for $T\Tbar$ perturbations of conformal field theories,  based mainly on factorization of the  CDD factor, was not derived from first principles,  but rather was guided by
the convergence of various integrals in the TBA equations, and comparison with different exact methods,  such as those based on the Burgers equation. 
 It would be desirable to have an alternative derivation of these equations, which may require some more scrutiny  of massless scattering.  
 
Secondly,  
the values of $\cstarUV$ we proposed  for the singular flows  for $c_{IR} >0$ are subject to the interpretation 
of the unconventional RG flow.   If the singularity in the UV unequivocally prevents flowing to arbitrarily short distances,  then the theory would formally appear to be UV
 incomplete in the usual sense of a local quantum field theory.  
 \footnote{Note added: ~In subsequent work we showed that it is possible to complete the flow to the UV by adding additional $T\Tbar$-like operators based on higher integrals of motion \cite{ALcompletion}.}.
 Our analysis  at least provides   a partial UV completion in this case.  
On the other hand, for $c_{IR} \leq 0$, we did prove that the UV theory is indeed complete.
  At best, physically there  truly is a shortest possible distance, and we have proposed the correct interpretation of the flow.     However the precise nature of 
 this UV theory at the singularity,  which is not a CFT,   in particular its field content and lagrangian,  is not entirely  predicted by the TBA.     However we did obtain one property of such a theory in Section VIB,  
 namely the scaling dimension of the operator that perturbs this UV theory toward the IR fixed point.  
 Also,  we suggested that the singularity signifies a tachyonic vacuum instability.   If the latter is correct,  then the theory is in principle UV complete, except that we don't know yet precisely what this theory is.   On the other hand,  as discussed above the minimal distance $R_*$ is consistent with a non-local string theory at short distances.

 The third  aspect that deserves further study concerns the possible connection of Section VII with the Liouville theory,  if any.   Our construction builds on the correct 
 $c_{IR}$ by  simply introducing a chemical potential,  but it is unclear if this is the unique manner in which to do this within the framework of the TBA.   In any case,  at the very least 
 we described a correct manner in which to vary $c_{IR}$ regardless of any possible connection with Liouville.   
 Perhaps a check of the excited state energies could confirm or rule out  such a connection.

\section*{Acknowledgments}

We wish to thank Dnyanesh Kulkarni and Giuseppe Mussardo for discussions.


\begin{thebibliography}{99}


\bibitem{ZTT}  A. B. Zamolodchikov, {\it Expectation value of composite field T anti-T in two-dimensional quantum field theory}, hep-th/0401146.

\bibitem{Tateo1}
A Cavagli\`a, S Negro, I.M. Szecsenyi, R Tateo, 
{\it $T\Tbar$  -deformed 2D quantum field theories}, JHEP 10 (2016) 112,
arXiv:1608.05534 [hep-th].

\bibitem{SmirnovZ}  F. A. Smirnov and A. B. Zamolodchikov, {\it On the  space of integrable quantum field theories}, Nucl. Phys. B 915 (2017) 363, [arXiv:1608.05499].

\bibitem{Verlinde}
L McGough, M Mezei, H Verlinde,  
{\it Moving the CFT into the bulk with $T\Tbar$},   
JHEP 10 (2018) 1, 
arXiv:1611.03470 [hep-th].


\bibitem{Dubovsky} 
S. Dubovsky, V. Gorbenko, and M. Mirbabayi, {\it Asymptotic fragility, near AdS2 holography and $T\Tbar$, } JHEP 09 (2017) 136, arXiv:1706.06604 [hep-th].

\bibitem{Rosenhaus}
Vladimir Rosenhaus, Michael Smolkin,
{\it Integrability and Renormalization under $T\Tbar$}, 
Phys. Rev. D 102, 065009 (2020), 
	arXiv:1909.02640 [hep-th]

\bibitem{Tateo2}
R. Conti, L. Iannella, S. Negro, and R. Tateo, {\it Generalised Born-Infeld models, Lax operators and the $T\Tbar$ perturbation,} JHEP 11 (2018) 007, arXiv:1806.11515 [hep-th].


\bibitem{Cardy}
J. Cardy, {\it The $T\Tbar$ deformation of quantum field theory as random geometry,} JHEP 10 (2018) 186, arXiv:1801.06895 [hep-th].


\bibitem{Dubovsky2}
S. Dubovsky, V. Gorbenko, and G. Hernandez-Chifflet, {\it $T\Tbar$ partition function from topological gravity,} JHEP 09 (2018) 158, arXiv:1805.07386 [hep-th].


\bibitem{Hartman}
T. Hartman, J. Kruthoff, E. Shaghoulian, and A. Tajdini, {\it Holography at finite cutoff with a $T^2$ deformation,} ~JHEP 03 (2019) 004, arXiv:1807.11401 [hep-th].

\bibitem{Tateo3}
R. Conti, S. Negro, and R. Tateo, {\it The $T\Tbar$ perturbation and its geometric interpretation,} JHEP 02 (2019) 085, arXiv:1809.09593 [hep-th].


\bibitem{Teitel}
 C. Teitelboim, {\it Gravitation and Hamiltonian Structure in Two Space-Time
Dimensions,}  Phys. Lett. B126 (1983) 41.

\bibitem{Jackiw} 
R. Jackiw, {\it Lower Dimensional Gravity,}  Nucl. Phys. B252 (1985) 343.


\bibitem{Kutasov}
S. Chakraborty,  A.  Giveon, N. Itzhaki and D. Kutasov,
{\it Entanglement beyond AdS},
arXiv:1805.06286 [hep-th]. 

\bibitem{Chakraborty}
S.  Chakraborty and A. Hashimoto,
{\it Thermodynamics of  $T\Tbar$ , $J\Tbar$ , $T\bar{J}$ deformed conformal field theories}, 
arXiv:2006.10271 [hep-th].


\bibitem{Frolov}
S. Frolov, {\it $T\Tbar$ deformation and the light-cone gauge,}  arXiv:1905.07946 [hep-th].

\bibitem{Oku}  S. Okumura and K. Yoshida, 
{\it $T\Tbar$ perturbation and Liouville gravity},
Nucl. Phys. B957 (2020) 115083.



\bibitem{Jiang}  Y. Jiang, 
{\it Lectures on solvable irrelevant deformations of 2d quantum field theory}, 
arXiv:1904.13376 [hep-th]. 

\bibitem{ZTBA}  
Al. B.  Zamolodchikov, {\it Thermodynamic Bethe Ansatz in relativistic models: scaling 3-state Potts nd Lee-Yang models}, 
Nucl.  Phys.  B342 (1990) 695.

\bibitem{KM}  T. Klassen and E. Melzer,
{\it The thermodynamics of purely elastic scattering theories and conformal perturbation theory},
Nucl. Phys. B350 (1991) 635.

\bibitem{DubovskyTBA}  S. Dubovsky, R. Flauger, and V. Gorbenko,
{\it Solving the Simplest Theory of Quantum Gravity},
JHEP 1209 (2012) 133,  arXiv:1205.6805 [hep-th]. 

\bibitem{Tateo0}  M. Caselle, D. Fioravanti, F. Gliozzi and R. Tateo, 
{\it Quantisation of the effective string with TBA},   JHEP 07  (2013) 71, 
arXiv:1305.1278 [hep-th].  

\bibitem{ZZ}   A. B. Zamolodchikov and    Al. B. Zamolodchikov,
{\it  Massless factorized scattering
and sigma models with topological terms,} 
Nucl. Phys. B379 (1992) 602.

\bibitem{Fendley}    P. Fendley, H. Saleur, Al.B. Zamolodchikov, {\it Massless flows II: the exact S-matrix approach}, Int. J. Mod. Phys. A 8
(1993) 5751,
arXiv:hep-th/930405.

\bibitem{MussardoSimon}  G. Mussardo and P. Simon, 
{\it Bosonic-type S-matrix, vacuum instability, and CDD ambiguity},
Nucl.Phys. B578 (2000) 527, 
[arXiv:hep-th/9903072] .


\bibitem{ctheorem}  A. B. Zamolodchikov, 
{\it Irreversibility of the flux of the renormalization group in a 2D field theory},
Pis'ma Eksp. Teor. Fiz. {\bf 43} (1986) 565. 

\bibitem{DHoker}   E. D'Hoker,  D. Z. Freedman, and R. Jackiw,
{\it  SO(2,1) quantization of the Liouville theory},
Phys. Rev. D28 (1983) 2583.

\bibitem{Yoneya}  T. Yoneya,  
{\it  Triviality of the S-matrix in the quantum Liouville field theory,}
Phys. Lett. B  (1984) 111.

\bibitem{Teschner} J. Teschner,
{\it Liouville theory revisited}, 
Class. Quantum Grav. {\bf 18} R153,
arXiv:hep-th/0104158.

\bibitem{LecMuss}  A. LeClair and G. Mussardo,  
{\it Finite Temperature Correlation Functions in Integrable QFT}, 
 Nucl.Phys. B552 (1999) 624,  
 arXiv:hep-th/9902075.
 
 \bibitem{ALcompletion}  A. LeClair,
 {\it $T\Tbar$ deformation of the Ising model and its ultraviolet completion},
 arXiv:2107.02230 [hep-th].   
 
 
 \bibitem{Fractional} D. Bernard and A. LeClair,
 {\it The fractional supersymmetric sine-Gordon models,}
 Phys. Lett. B 247 (1990) 309. 
 
\bibitem{BL}
D. Bernard and A. LeClair, 
{\it Differential Equations for Sine-Gordon Correlation Functions at the Free Fermion Point}, 
Nucl.Phys. B426 (1994) 534, 
arXiv:hep-th/9402144;  Erratum-ibid. B498  (1997). 


\end{thebibliography}
\end{document}